\documentclass[twocolumn,english,showpacs,aps,prl]{revtex4}
\usepackage[T1]{fontenc}
\usepackage[latin1]{inputenc}
\usepackage{amsmath}
\usepackage{graphicx}
\usepackage{amssymb}

\makeatletter
\usepackage{graphicx}
\usepackage{epsfig}
\usepackage{multirow}
\usepackage{dcolumn}
\usepackage{bm}
\usepackage{subfigure}

\def\<{\langle}
\def\>{\rangle}

\usepackage{babel}
\makeatother
\begin{document}

\title{The low energy spectrum of finite size metallic SWNTs}

\author{Leonhard Mayrhofer and Milena Grifoni}

\affiliation{Theoretische Physik, Universit\"{a}t Regensburg, 93040 Germany}

\date{\today{}}

\begin{abstract}
The electronic spectrum of metallic finite-size single-wall carbon
nanotubes at low energies is derived. It is based on a tight-binding
description for the interacting $p_{z}$ electrons. Not only the forward
scattering parts of the Coulomb interaction, which are diagonalized
by bosonization, are considered, but also all other processes becoming
relevant for small diameter tubes. As a consequence of the substructure
of the underlying lattice, a spin 1 triplet is found as ground state
if the exchange splitting is larger than the branch mismatch, a spin
0 singlet otherwise. Moreover the excitation spectrum is calculated.
\end{abstract}


\maketitle
Single-walled carbon nanotubes (SWNTs) are one of the most prominent
examples for the realization of 1D electronic systems with orbital
degeneracy in nature. A proper description of the low energy regime
has to take into account the Coulomb interaction between the electrons
and the corresponding correlation effects. Luttinger Liquid behaviour,
leading to power-law dependence of various transport quantities, has
been predicted theoretically for metallic SWNTs of infinite length
\cite{Egger1997,Odintsov1999} and was confirmed experimentally \cite{Bockrath1999,Postma2001}.
Considering tubes of finite length, Kane et al. \cite{Kane1997} have
derived the discrete energy spectrum of the collective spin and charge
excitations and its dependence on the forward scattering part of the
Coulomb interaction. Moreover, for SWNTs a charging energy of the
order of the level spacing is found, leading to the observation of
Coulomb blockade in SWNT quantum dot devices \cite{Tans1997,Cobden2002,Liang2002,Sapmaz2005,Moriyama2005}.
The accompanying two- or fourfold periodicity of the Coulomb diamond
size can be understood by including the spin and band degrees of freedom.
A theory of transport through SWNT quantum dots incorporating the
mentioned forward scattering interaction processes was worked out
in \cite{Mayrhofer2006}. As discussed below, the restriction to forward
scattering terms is justified for large diameter tubes only, whereas
the remaining interaction processes become more and more important
for decreasing tube diameters. They lead to exchange effects and a
modification of the excitation spectrum. Within a meanfield approach
Oreg et al. \cite{Oreg2000} predicted exchange effects favouring
spin alignment. The size of the exchange splitting on the SWNT ground
states was measured by recent experiments \cite{Liang2002,Sapmaz2005,Moriyama2005}. 

A fundamental question regards the nature of the ground state and
excited spectrum of correlated 1D systems. In a milestone theorem
Lieb and Mattis \cite{Lieb-Mattis1962} demonstrated that for a 1D
Hubbard model with nearest neighbour hopping, the ground state must
have spin $0$ or $1/2$. They left open the question for systems
with orbital degeneracy. This is the case for SWNTs due to the substructure
of the underlying honeycomb lattice.

In this work we go beyond the mean field approach and derive the electronic
structure of metallic SWNTs in the low energy regime from a microscopic
model. Beside the long ranged forward scattering terms of the Coulomb
interaction, which are exactly diagonalized by bosonization, we take
into account \textit{all} other processes becoming relevant for small
diameter SWNTs away from half-filling. Because of scattering processes
involving the orbital degree of freedom, we predict a spin $1$ triplet
as ground state, if the exchange energy exceeds the energy mismatch
between the two electron branches and if $4m+2$ electrons occupy
the SWNT. Moreover the excitation spectra are calculated. The large
degeneracies of the discrete energy levels, as obtained by only including
forward scattering terms, are partly lifted and the spectra become
quasi-continuous when going to higher energies.

\paragraph*{The low energy Hamiltonian of metallic finite size SWNTs}

Without loss of generality we focus on armchair SWNTs \cite{Odintsov1999_2}.
As basis of our future discussion let us recall the minimal model
Hamiltonian to describe finite size armchair SWNTs at low energies
derived in \cite{Mayrhofer2006}. It is based on $p_{z}$ electrons
localized on the graphene honeycomb lattice which contains two carbon
atoms, $p=\pm$, per unit cell. Let the SWNT axis be along the $x$
direction and ignore for the moment interactions. Then, by imposing
periodic boundary conditions along the circumference and open ones
along the tube length, eigenfunctions are standing waves $\varphi_{r\kappa}(\vec{r})$
with the branch or pseudo-spin index $r=\pm$. The wave number $\kappa$
describes the slowly varying oscillations of $\varphi_{r\kappa}(\vec{r})$.
The finite tube length $L$ leads to the quantization condition $\kappa=\pi(m_{\kappa}+\Delta)/L,$
$m_{\kappa}\in\mathbb{Z},\,\left|\Delta\right|\le1/2$. The parameter
$\Delta$ is responsible for a possible energy mismatch between the
branches $r=\pm$; its value depends on the length and on the type
of the considered SWNT \cite{Jiang2002}. Explicitly, $\varphi_{r\kappa}(\vec{r})$
can be decomposed into its contributions from the sublattices $p=\pm$,
$\varphi_{r\kappa}(\vec{r})=\frac{1}{\sqrt{2}}\sum_{p=\pm}f_{pr}\sum_{F=\pm K_{0}}\mathrm{sgn}(F)e^{i\mathrm{sgn}(F)\kappa x}\varphi_{pF}(\vec{r}),$
where $f_{+\, r}=1/\sqrt{2}$ and $f_{-\, r}=-r/\sqrt{2}$. The functions
$\varphi_{pF}(\vec{r})$ describe fast oscillating Bloch waves at
the two independent Fermi points $F=\pm K_{0}$ of the honeycomb lattice,
$\varphi_{pF}(\vec{r})=N_{L}^{-1/2}\sum_{\vec{R}}e^{iFR_{x}}\chi(\vec{r}-\vec{R}-\vec{\tau}_{p}),$
where $\chi(\vec{r}-\vec{R}-\vec{\tau}_{p})$ is the $p_{z}$ orbital
at lattice site $\vec{R}$ on sublattice $p$ and $N_{L}$ is the
total number of lattice sites. The non-interacting Hamiltonian then
reads\[
H_{0}=\hbar v_{F}\sum_{r\sigma}r\sum_{\kappa}\kappa c_{r\sigma\kappa}^{\dagger}c_{r\sigma\kappa},\]
where $v_{F}\approx8.1\cdot10^{5}\textrm{ m/s}$ is the Fermi velocity
of graphene and $c_{r\sigma\kappa}$ annihilates an electron in the
state $\left|\varphi_{r\kappa}\right\rangle \left|\sigma\right\rangle $.
Introducing the slowly varying 1D operators $\psi_{r\sigma F}(x)$
\cite{Mayrhofer2006} defined along the tube axis,\begin{equation}
\psi_{r\sigma F}(x)=\frac{1}{\sqrt{2L}}\sum_{\kappa}e^{i\mathrm{sgn}(F)\kappa x}c_{r\sigma\kappa},\label{eq:1Deop_c}\end{equation}
 and integrating over the coordinates perpendicular to the tube axis,
the interaction part of the Hamiltonian becomes effectively one dimensional.
We find\begin{multline}
V=\frac{1}{2}\sum_{\sigma\sigma'}\sum_{\{[r],[F]\}}\prod_{i=1}^{4}\mathrm{sgn}(F_{i})\int\int dx\, dx'U_{[r][F]}(x,x')\\
\times\psi_{r_{1}\sigma F_{1}}^{\dagger}(x)\psi_{r_{2}\sigma'F_{2}}^{\dagger}(x')\psi_{r_{3}\sigma'F_{3}}(x')\psi_{r_{4}\sigma F_{4}}(x).\label{eq:V1D}\end{multline}
Here $\sum_{\{[r],[F]\}}$ denotes the sum over all quadruples $[r]=(r_{1},r_{2},r_{3},r_{4})$
and $[F]=(F_{1},F_{2},F_{3},F_{4}).$ Assuming that the wave functions
$\varphi_{p,F}(\vec{r})$ and $\varphi_{-p,F}(\vec{r})$ do not overlap,
the effective 1D Coulomb interaction potential \begin{multline}
U_{[r][F]}(x,x')=\frac{1}{4}\left[U_{[F]}^{intra}(x,x')\left(1+r_{1}r_{2}r_{3}r_{4}\right)\right.+\\
\left.U_{[F]}^{inter}(x,x')\left(r_{2}r_{3}+r_{1}r_{4}\right)\right],\label{eq:U1D_rF}\end{multline}
 can be separated into an interaction for electrons on the same (intra)
and on different sublattices (inter), where\begin{multline*}
U_{[F]}^{intra/inter}(x,x')=L^{2}\int\int d^{2}r_{\perp}d^{2}r'_{\perp}\\
\times\varphi_{pF_{1}}^{*}(\vec{r})\varphi_{\pm pF_{2}}^{*}(\vec{r}\,')\varphi_{\pm pF_{3}}(\vec{r}\,')\varphi_{pF_{4}}(\vec{r})U(\vec{r}-\vec{r}\,'),\end{multline*}
and $U(\vec{r}-\vec{r}\,')$ is the Coulomb potential. For the actual
calculations we model $U(\vec{r}-\vec{r}\,')$ by the so called Ohno
potential. Measuring distances in units of $\textrm{Å}$ and energy
in $\mathrm{eV}$, it is given by $U(\vec{r}-\vec{r}\,')=U_{0}/\sqrt{1+\left(U_{0}\epsilon\left|\vec{r}-\vec{r}\,'\right|/14.397\right)^{2}}\,\mathrm{eV}$
\cite{Barford}. A reasonable choice is $U_{0}=15$ eV \cite{Oreg2000}.
The dielectric constant is given by $\epsilon\approx1.4-2.4$ \cite{Egger1997}.

\paragraph{The relevant scattering processes}

To proceed, it is convenient to introduce the notion of forward ($f$)-,
back ($b$)-, and Umklapp ($u$)- scattering for an arbitrary index
quadruple $[I]$ associated to the four electron operators in (\ref{eq:V1D}).
With $S_{I}$ being the type of scattering process for the quantity
$I$, we denote a quadruple $[I,\pm I,\pm I,I]$ by $[I]_{S_{I}=f^{\pm}}$,
while $[I]_{b}$ is equivalent to $[I,-I,I,-I].$ Finally Umklapp
scattering means $[I]_{u}=[I,I,-I,-I].$ As we will show in the sequel
the interaction part $V$ of the Hamiltonian is of the form $V=\sum_{S_{r}=f,b,u}\sum_{S_{F}=f,b}\sum_{S_{\sigma}=f}V_{S_{r}S_{F}S_{\sigma}},$
where \begin{multline}
V_{S_{r}S_{F}S_{\sigma}}:=\frac{1}{2}\sum_{\left\{ [r]_{S_{r}},[F]_{S_{F}},[\sigma]_{S_{\sigma}}\right\} }\int\int dx\, dx'U_{[r][F]}(x,x')\\
\times\psi_{r_{1}\sigma F_{1}}^{\dagger}(x)\psi_{r_{2}\sigma'F_{2}}^{\dagger}(x')\psi_{r_{3}\sigma'F_{3}}(x')\psi_{r_{4}\sigma F_{4}}(x).\label{eq:nonrhorho1}\end{multline}

We start with the $S_{r}$ scattering types. From (\ref{eq:U1D_rF})
it is evident, that the effective interaction potential is only nonzero
for $r_{2}r_{3}=r_{1}r_{4}.$ Hence we have to distinguish between
the following two cases,\[
a)\, r_{1}=r_{4},\, r_{2}=r_{3}\textrm{ and }b)\, r_{1}=-r_{4},\, r_{2}=-r_{3}.\]
 In case $a)$ we have $S_{r}=f$ and $U_{[r]_{f}[F]}(x,x')$ is the
sum of intra- and inter- lattice interaction. Case $b)$ comprises
$S_{r}=b$ and $S_{r}=u$, with $U_{[r]_{b/u}[F]}(x,x')$ being the
difference of the two types of sublattice interactions. Now we look
at the allowed $S_{F}$ processes. Although not dealing with an infinite
system, after the integrations along the tube axis in (\ref{eq:V1D}),
only terms with $F_{1}+F_{2}-F_{3}-F_{4}=0,$ i.e. the $S_{F}=f$
and $S_{F}=b$ processes, dominate as a consequence of the approximate
conservation of quasi momentum. All other processes, including $S_{F}=u$,
have very small amplitudes. Finally, regarding the spin index, only
$S_{\sigma}=f$ terms are allowed, since the Coulomb interaction is
spin independent.

Additionally, away from half filling, only terms with $r_{1}F_{1}+r_{2}F_{2}-r_{3}F_{3}-r_{4}F_{4}=0$
are relevant in (\ref{eq:V1D}), due to conservation of the quasi
momentum, arising from the slow oscillations of the 1D electron operators. 

Since $U_{[F]}^{intra}$ and $U_{[F]}^{inter}$ differ only at the
length scale of the lattice spacing \cite{Egger1997}, the interaction
potential for case $b)$ ($\rightarrow S_{r}=b,\, u$) is generally
short ranged compared to the slowly varying electron operators $\psi_{r\sigma F}$,
whereas in case $a)$ ($\rightarrow S_{r}=f$), this is only true
for the $S_{F}=b$ terms. Hence for $S_{r}=b,\, u$ or $S_{F}=b$
the corresponding interactions become effectively local: \begin{multline}
V_{S_{r}S_{F}S_{\sigma}}:=Lu^{S_{r}\, S_{F}}\sum_{\left\{ [r]_{S_{r}},[F]_{S_{F}},[\sigma]_{FS}\right\} }\\
\times\int_{0}^{L}dx\psi_{r_{1}\sigma F_{1}}^{\dagger}(x)\psi_{r_{2}\sigma'F_{2}}^{\dagger}(x)\psi_{r_{3}\sigma'F_{3}}(x)\psi_{r_{4}\sigma F_{4}}(x),\label{eq:nonrhorhob}\end{multline}
where the coupling constants are given by $u^{S_{r}\, S_{F}}=1/\left(2L^{2}\right)\int\int dx\, dx'U_{[r]_{S_{r}},[F]_{S_{F}}}(x,x').$
It holds $u^{b\, S_{F}}=u^{u\, S_{F}}=:u^{\Delta\, S_{F}}$ and we
define $u^{+}:=u^{f\, b}.$ The ratio $u^{S_{r}\, S_{F}}/\varepsilon_{0}$
is independent of $L$ but scales like $1/d$ where $d$ is the tube
diameter, such that these processes become negligible for large diameter
tubes. Numerically we find $u^{+}\approx u^{\Delta\, b}=0.22\,[0.28]\frac{\varepsilon_{0}}{d}\textrm{Å}$
and $u^{\Delta\, f}=0.14\,[0.22]\frac{\varepsilon_{0}}{d}\textrm{Å}$
for $\epsilon=1.4\,[2.4]$.

\paragraph{Density-density interactions}

In the next step we introduce the quantities $V_{\rho\rho}$ and $V_{\mathrm{n\rho\rho}}$
that collect all interaction processes that are of density-density
and non-density-density form, respectively, such that in total $H=H_{0}+V_{\rho\rho}+V_{\mathrm{n}\rho\rho}$
. Since the short ranged interactions are treated as local, we obtain
\[
V_{\rho\rho}:=V_{f\, f\, f}+V_{f^{+}\, b\, f^{+}}+V_{b\, f^{+}/b\, f^{+}},\]
where the dominating part of the interaction away from half filling
is the long-ranged term $V_{f\, f\, f}$. Using bosonization techniques
\cite{Delft1998}, the Hamiltonian $H_{0}+V_{\rho\rho}$ can be diagonalized
and we find\begin{multline}
H_{0}+V_{\rho\rho}=\sum_{j\delta}\sum_{q>0}\varepsilon_{j\delta q}a_{j\delta q}^{\dagger}a_{j\delta q}+\frac{1}{2}E_{c}\mathcal{N}_{c}^{2}\\
+\frac{1}{2}\sum_{r\sigma}\mathcal{N}_{r\sigma}\left[\mathcal{N}_{r\sigma}\left(\varepsilon_{0}-u^{+}\right)+r\varepsilon_{\Delta}-\frac{J}{2}\mathcal{N}_{-r\sigma}\right],\label{eq:Hrhorho}\end{multline}
where we have defined $J:=2(u^{\Delta\, f}+u^{\Delta\, b})$. The
first term describes discrete excitations created/annihilated by the
bosonic operators $a_{j\delta q}/a_{j\delta q}^{\dagger}$. The four
channels $j\delta=c+,c-,s+,s+$ are related to total and relative
(with respect to the $r$ index) charge and spin excitations. With
the level spacing of the free electrons, $\varepsilon_{0}:=\hbar v_{F}\frac{\pi}{L}$,
and $\varepsilon_{0q}:=\varepsilon_{0}n_{q},\, q=n_{q}\frac{\pi}{L}$
the relations $\varepsilon_{c+q}=\varepsilon_{0q}\sqrt{1+8W_{q}/\varepsilon_{0}}$,
$\varepsilon_{s/c-q}=\varepsilon_{0q}(1-u^{\Delta\, b}/\varepsilon_{0})$
and $\varepsilon_{s+q}=\varepsilon_{0q}(1+u^{\Delta\, b}/\varepsilon_{0})$
hold. Due to the dominating $V_{f\, f\, f}$ contribution, the ratio
$g_{q}:=\varepsilon_{0q}/\varepsilon_{c+q}$ is strongly reduced for
small $q$ ($g_{q}\approx0.2$) and for large $q$, $g_{q}$ tends
to $1$ \cite{Mayrhofer2006}. Small corrections due to the coupling
constants $u^{+}$ and $u^{\Delta\, f}$ have been omitted. Finally,
\begin{multline*}
W_{q}=\frac{1}{L^{2}}\int_{0}^{L}dx\int_{0}^{L}dx'\cos(qx)\cos(qx')U_{[r]_{f}[F]_{f}}(x,x').\end{multline*}
 The remaining terms in (\ref{eq:Hrhorho}) are fermionic contributions,
accounting for the energy cost of changing the number of electrons
in the different branches $(r\sigma)$. The operators $\mathcal{N}_{r\sigma}$
count the electrons in $(r\sigma)$ and $\mathcal{N}_{c}=\sum_{r\sigma}\mathcal{N}_{r\sigma}$.
The single summands account for (in the order of appearance) the Coulomb
charging energy $E_{c}=W_{0},$ the shell filling energy (because
of Pauli's principle), a possible energy mismatch $\varepsilon_{\Delta}=\mathrm{sgn}(\Delta)\varepsilon_{0}\min(2\left|\Delta\right|,2\left|\Delta\right|-1)$
between the $r$ branches if $\left|\Delta\right|\neq0,1/2$, and
a favourable spin alignment of electrons with different $r$ due to
$V_{b\, f^{+}/b\, f^{+}}$. Note also that the shell filling energy
is modified by the attractive contribution $-u^{+}$ due to $V_{f^{+}\, b\, f^{+}}$.
The eigenstates of $H_{0}+V_{\rho\rho}$ are $|\vec{N},\vec{m}\left.\right\rangle :=\prod_{j\delta q}\left(a_{j\delta q}^{\dagger}\right)^{m_{n\delta q}}/\sqrt{m_{j\delta q}!}|\vec{N},0\left.\right\rangle ,$
where $|\vec{N},0\left.\right\rangle $ has no bosonic excitation
and $\vec{N}=(N_{-\uparrow},N_{-\downarrow},N_{+\uparrow},N_{+\downarrow})$
defines the number of electrons in each of the branches $(r\sigma)$.

\paragraph{Non-density-density processes}

In the following we concentrate on the situation away from half-filling,
where the non-density-density processes $V_{\mathrm{n}\rho\rho}$
can be treated as a small perturbation to the Hamiltonian $H_{0}+V_{\rho\rho}$.
We calculate the low energy spectrum and the corresponding eigenstates
of the full Hamiltonian $H_{0}+V_{\rho\rho}+V_{\mathrm{n}\rho\rho}$
by evaluating the matrix elements $\left\langle \right.\vec{N},\vec{m}|V_{\mathrm{n}\rho\rho}|\vec{N}',\vec{m}'\left.\right\rangle $
and by truncating the Hilbert space at a certain eigenenergy of $H_{0}+V_{\rho\rho}.$
Near half-filling the strength of $V_{\mathrm{n}\rho\rho}$ is highly
enhanced and the truncation procedure therefore questionable. We obtain\begin{equation}
V_{\mathrm{n}\rho\rho}=V_{f^{+}\, b\, f^{-}}+V_{b\, f^{+}/b\, f^{-}}+V_{u\, f^{-}/b\, f}.\label{eq:Vnrr_expl}\end{equation}

\paragraph{Low energy spectrum}

Our truncation scheme to find the low energy spectrum is to only retain
the energetically lowest lying states of $H_{0}+V_{\rho\rho}$ which
have no bosonic excitations. Using (\ref{eq:1Deop_c}) and (\ref{eq:nonrhorhob}),
we get for the matrix elements of the contributions on the r.h.s of
(\ref{eq:Vnrr_expl}), 

\begin{multline}
\left\langle \vec{N},0\left|V_{S_{r}S_{F}S_{\sigma}}\right|\vec{N}',0\right\rangle =\frac{1}{4}u^{S_{r}\, S_{F}}\sum_{\left\{ [r]_{S_{r}},[F]_{S_{F}},[\sigma]_{FS}\right\} }\\
\times\sum_{\kappa_{1},\dots,\kappa_{4}}\left\langle \vec{N},0\left|c_{r_{1}\sigma\kappa_{1}}^{\dagger}c_{r_{2}\sigma'\kappa_{2}}^{\dagger}c_{r_{3}\sigma'\kappa_{3}}c_{r_{4}\sigma\kappa_{4}}\right|\vec{N}',0\right\rangle \\
\times\delta_{\sum_{i=1}^{2}sgn(F_{i})\kappa_{i}-\sum_{i=3}^{4}sgn(F_{i})\kappa_{i},0},\label{eq:V_nrr_gstates}\end{multline}
where the Kronecker-$\delta$ results from the integration along the
tube axis in (\ref{eq:nonrhorhob}). Note that the states $|\vec{N},0\left.\right\rangle $
are eigenstates of $H_{0}+V_{\rho\rho}$ and not of $H_{0}$ alone.
Hence the evaluation of (\ref{eq:V_nrr_gstates}) in general is not
straightforward. But for the processes relevant away from half-filling
we get the same result and physical insight if we consider for the
calculation of (\ref{eq:V_nrr_gstates}) $|\vec{N},0\left.\right\rangle $
as the Fermi sea filled up with $N_{r\sigma}$ \textit{noninteracting}
electrons in the branch $r\sigma$, i. e. as an eigenstate of $H_{0}$,
as we will discuss elsewhere \cite{Mayrhofer2007}.

\begin{figure}
\begin{center}\includegraphics[%
  width=0.87\columnwidth,
  keepaspectratio]{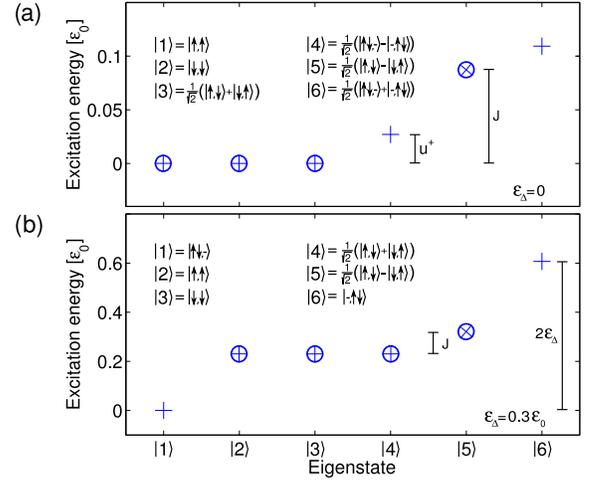}\end{center}

\caption{\label{cap:Ground-states}Low energy spectrum of a $(6,6)$ SWNT
for $N_{c}=4m+2$. (a) For $\varepsilon_{\Delta}=0$ the ground state
is formed by the spin $1$ triplet $(\rightarrow\oplus)$. The states
$\left|\uparrow\downarrow,-\right\rangle $ and $\left|-,\uparrow\downarrow\right\rangle $
mix. (b) For $\varepsilon_{\Delta}\gg J/2$ the ground state is given
by the spin $0$ state $\left|\uparrow\downarrow,-\right\rangle .$
The spin $0$ singlet is indicated by $\otimes$. The interaction
parameters are $J=0.09\varepsilon_{0}$, $u^{+}=0.03\varepsilon_{0}$.}
\end{figure}

In the following we focus on the case $N_{c}=4m+2$. As truncated
basis we use the states $|\vec{N},0\left.\right\rangle $ with $\vec{N}=(m+1,m+1,m,m)$
and permutations. In the following we denote $\left|(m+1,m+1,m,m),0\right\rangle $
by $\left|\uparrow\downarrow,-\right\rangle $ etc. Using (\ref{eq:Hrhorho})
and (\ref{eq:V_nrr_gstates}) the interacting Hamiltonian, restricted
to the states $\left|\uparrow,\uparrow\right\rangle $, $\left|\downarrow,\downarrow\right\rangle $,
$\left|\uparrow,\downarrow\right\rangle $, $\left|\downarrow,\uparrow\right\rangle $,
$\left|\uparrow\downarrow,-\right\rangle $ and $\left|-,\uparrow\downarrow\right\rangle $,
is\begin{multline}
H=E_{0,4m+2}+\\
\left(\begin{array}{ccccccc}
-\frac{J}{2} &  &  &  &  &  & 0\\
 &  & -\frac{J}{2}\\
 &  &  & 0 & -\frac{J}{2}\\
 &  &  & -\frac{J}{2} & 0\\
 &  &  &  &  & u^{+}-\varepsilon_{\Delta} & \frac{J}{2}\\
0 &  &  &  &  & \frac{J}{2} & u^{+}+\varepsilon_{\Delta}\end{array}\right),\label{eq:H_4mp2}\end{multline}
with $E_{0,4m+2}=\frac{1}{2}E_{c}N_{c}^{2}+(2m^{2}+2m+1)\left(\varepsilon_{0}-u^{+}\right)-J(m^{2}+m)+2u^{+}m$.
Diagonalizing $H$, we find that its eigenstates are given by the
spin $1$ triplet $\left|\uparrow,\uparrow\right\rangle $, $\left|\downarrow,\downarrow\right\rangle $,
$1/\sqrt{2}\left(\left|\uparrow,\downarrow\right\rangle +\left|\downarrow,\uparrow\right\rangle \right)$,
the spin $0$ singlet $1/\sqrt{2}\left(\left|\uparrow,\downarrow\right\rangle -\left|\downarrow,\uparrow\right\rangle \right)$
and the two states $(c_{1/2}^{2}+1)^{-1/2}\left(c_{1/2}\left|\uparrow\downarrow,-\right\rangle \pm\left|-,\uparrow\downarrow\right\rangle \right)$,
where $c_{1/2}=\frac{J}{2}/\left(\sqrt{\varepsilon_{\Delta}^{2}+(J/2)^{2}}\pm\varepsilon_{\Delta}\right).$
Relatively to $E_{0,4m+2}$, the eigenenergies are $-J/2$ for the
triplet states, $J/2$ for the singlet state and $u^{+}\pm\sqrt{\varepsilon_{\Delta}^{2}+(J/2)^{2}}$
for the remaining two states. Thus, under the condition $J/2>\sqrt{\varepsilon_{\Delta}^{2}+(J/2)^{2}}-u^{+},$
i.e. for a small band mismatch $\varepsilon_{\Delta}\lesssim J/2$,
the ground state is threefold degenerate and formed by the spin $1$
triplet, otherwise by $(c_{2}^{2}+1)^{-1/2}\left(c_{2}\left|\uparrow\downarrow,-\right\rangle -\left|-,\uparrow\downarrow\right\rangle \right)$.
The low energy spectra for the two cases $\varepsilon_{\Delta}=0$
and $\varepsilon_{\Delta}\gg J/2$ are shown in Fig. \ref{cap:Ground-states}
for a (6,6) armchair nanotube (corresponding to $d=0.8$ nm) with
an assumed dielectric constant of $\epsilon=1.4$. The obtained values
of $J=0.09\varepsilon_{0}$ and $u^{+}=0.03\varepsilon_{0}$ are in
good agreement with the experiments \cite{Liang2002,Moriyama2005},
where nanotubes with $\varepsilon_{\Delta}\gg\frac{J}{2}$ were considered.
In accordance with the discussion above, Moriyama et al. \cite{Moriyama2005}
could identify the ground state to $N_{c}=4m+2$ as a spin $0$ singlet
and also a spin $1$ triplet was found. Not observed so far has been
the spin $1$ triplet as ground state and the mixing of $\left|\uparrow\downarrow,-\right\rangle $
and $\left|-,\uparrow\downarrow\right\rangle $, as predicted by our
calculation for nanotubes with $\varepsilon_{\Delta}\lesssim J/2$.
We emphasize that all the exchange splittings here result from non-forward
scattering processes with respect to the band index $r$. In the considerations
of Lieb and Mattis \cite{Lieb-Mattis1962} such an additional pseudo-spin
degree of freedom is missing and so we conjecture that this is the
reason why their theorem can not be applied in our case. The meanfield
result of Oreg et al. \cite{Oreg2000} for the ground state structure
to $N_{c}=4m+2$ essentially can be obtained by setting the off-diagonal
elements in (\ref{eq:H_4mp2}) to zero. Therefore if $\varepsilon_{\Delta}\gg\frac{J}{2}$,
the meanfield approach yields the same ground state spectrum as our
work, but with \textit{different degeneracies}: Instead of having
a threefold degeneracy of the spin $1$ triplet and no degeneracy
for the spin $0$ singlet, a twofold degeneracy of the states $\left|\uparrow,\uparrow\right\rangle $
, $\left|\downarrow,\downarrow\right\rangle $ and $\left|\uparrow,\downarrow\right\rangle $,
$\left|\downarrow,\uparrow\right\rangle $ respectively, is obtained.
Additionally for $\varepsilon_{\Delta}\lesssim J/2$, the meanfield
theory is not capable of predicting the mixing of the states $\left|\uparrow\downarrow,-\right\rangle $
and $\left|-,\uparrow\downarrow\right\rangle $ with the accompanying
exchange splitting. %
 The ground states of $H_{0}+V_{\rho\rho}$ to the charge states $N_{c}\neq4m+2$
do not mix via $V_{\mathrm{n}\rho\rho}$ and thus the corresponding
energies can be determined easily by using eqs. (\ref{eq:Hrhorho})
and (\ref{eq:V_nrr_gstates}) for $\vec{N}=\vec{N}'$.

\paragraph{Excitation spectrum}

\begin{figure}
\begin{center}\includegraphics[%
  width=0.85\columnwidth,
  keepaspectratio]{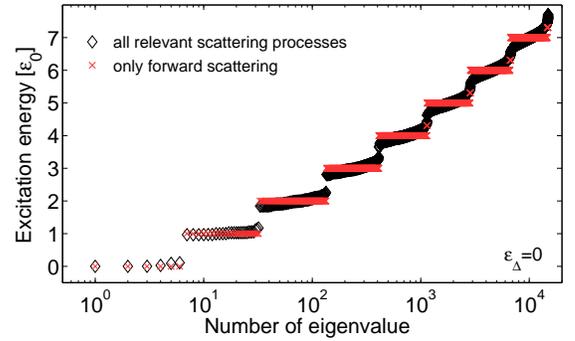}\end{center}

\caption{\label{cap:Excitation-spectrum-for}Excitation spectrum of a (6.6)
SWNT including only forward scattering processes $(\rightarrow\times)$
and for the total Hamiltonian $H=H_{0}+V_{\rho\rho}+V_{\mathrm{n}\rho\rho}$
$(\rightarrow\diamondsuit)$. Note the logarithmic scale of the $x$-axis.
The energy of the lowest $c+$ excitation is $4.3\,\varepsilon_{0}$.
All other interaction parameters are as for Fig. \ref{cap:Ground-states}.}
\end{figure}

So far no bosonic excitations were involved in our ground state examination.
In order to discuss the excitation spectrum of $H=H_{0}+V_{\rho\rho}+V_{\mathrm{n}\rho\rho}$,
the matrix elements of $\left\langle \right.\vec{N},\vec{m}|V_{\mathrm{n\rho\rho}}|\vec{N}',\vec{m}'\left.\right\rangle $
must be determined. The detailed calculation, based on bosonizaton
techniques, will be given in a longer article \cite{Mayrhofer2007}.
After truncating the Hilbert space at sufficiently high energies and
diagonalizing $H$ we obtain for $N_{c}=4m+2$ the spectrum as shown
in Fig. \ref{cap:Excitation-spectrum-for}. For comparison we also
show the spectrum of the {}``standard'' theory \cite{Egger1997,Kane1997,Mayrhofer2006}
as obtained by only retaining the dominating forward scattering processes
of $V_{\rho\rho}$. Most striking is the partial lifting of the huge
degeneracies and the formation of a quasi continuum at higher energies.

In conclusion we have derived a microscopic low energy theory for
finite size metallic SWNTs away from half filling, including non-density-density
interaction processes, which become relevant for small diameter tubes.
The ground state and excitation spectra have been determined. In particular,
we predict a spin $1$ triplet as ground state for $N_{c}=4m+2$,
if the energy mismatch $\varepsilon_{\Delta}$ between the different
pseudo-spin branches is much smaller than the exchange energy $J,$
a spin $0$ singlet otherwise. For $\varepsilon_{\Delta}\lesssim J/2$
we furthermore find that the pseudo-spin is not conserved and the
corresponding degeneracy is lifted. We notice that in \cite{Moriyama2005}
an energy mismatch $\varepsilon_{\Delta}\gg\frac{J}{2}$ was used
to fit the data and, in agreement with our theory, a singlet ground
state has been inferred from magnetic field measurements. Observation
of the triplet ground state is within experimental reach.

\end{document}